\documentclass[twocolumn,secnumarabic,amssymb, nobibnotes,aps,prl]{revtex4-1}
\usepackage{graphicx}
\usepackage{textcomp}
\usepackage{amsmath}
\usepackage{commath}
\usepackage{color}

\newcommand{\sg}[1]{}
\renewcommand{\sg}[1]{{\color{black}{#1}}} 
\newcommand{\kz}[1]{}
\renewcommand{\kz}[1]{{\color{black}{#1}}} 
\newcommand{\kzbis}[1]{}
\renewcommand{\kzbis}[1]{{\color{black}{#1}}} 
\newcommand{\wj}[1]{}
\renewcommand{\wj}[1]{{\color{black}{#1}}} 

\begin{document}
	\title{Tunable bandgaps and symmetry breaking in magneto-mechanical metastructures \\ inspired by multi-layer 2D materials}
	\author{Kuan Zhang}
	\email{Contributed equally to the work}
    \author{Weijian Jiao}
    \email{Contributed equally to the work}
	\author{Stefano Gonella}
	\email{sgonella@umn.edu}
	\affiliation{Department of Civil, Environmental, and Geo- Engineering\\ University of Minnesota, Minneapolis, MN 55455, USA\\}
	
	\begin{abstract}
    In this Letter, we introduce a paradigm to realize magneto-mechanical metastructures inspired by multi-layer 2D materials, such as graphene bilayers. The metastructures are intended to capture two aspects of their nanoscale counterparts. One is the multi-layer geometry, which is implemented by stacking hexagonal lattice sheets. The other is the landscape of weak inter-layer forces, which is mimicked by the interactions between pairs of magnets located at corresponding lattice sites on adjacent layers. We illustrate the potential of this paradigm through a three-layer prototype. The two rigid outer lattices serve as control layers, while the thin inner layer is free to experience flexural motion under the confining action of the magnetic forces exchanged with the outer ones, thus behaving as a lattice on elastic foundation. The inner layer is free to rotate relatively to the others, giving rise to a rich spectrum of inter-layer interaction patterns. Our objective is to determine how the dynamical response can be tuned by changing the twist angle between the layers. Specifically, we demonstrate experimentally that switching between different stacking patterns has profound consequences on the phonon landscape, opening and closing bandgaps in different frequency regimes. 
		\vspace{0.4cm}
	\end{abstract}
	
	\maketitle
	
	\kzbis{2D heterostructures are layered nanostructures consisting of stacks of 2D materials that are characterized by strong in-plane bonding and weak van der Waals interactions between the layers~\cite{Geim-Grigorieva_Heterostructures_Nature_2013}. A prototypical example is graphene bilayers, which consist of two stacked layers of graphene~\cite{Dai-et-al_Graphene-Bilayers_Nano-Lett_2016,Yoo-et-al_Graphene-Bilayers_Nat-Mat_2019,Andrei-Macdonald_Graphene-Bilayers_Nat-Mat_2020, Zhang-Tadmor_Moire_EML_2017,Zhang-Tadmor_Twisted-graphene_JMPS_2018}. Because of the inter-layer interaction, bilayers feature a more complex set of functionalities compared to their monolayer counterparts. For example, if the layers are free to undergo a relative twist, a Moir\'e interference pattern appears and special stacking configurations (AA, AB/BA and saddle-point (SP)) with distinctive mechanical characters can be identified. Additionally, twists at the so-called magic angles lead to a series of unique behaviors, including superconductivity, formation of insulating states, and electronic nematicity.} 
	
%
%
%
%


	The objective of this Letter is to introduce a class of multi-layer mechanical metastructures qualitatively inspired by the morphology and inter-layer mechanisms of graphene bilayers, and determine how their dynamical response can be tuned by the relative twist of the layers. In recent years, a plethora of elastic metastructural designs have been proposed to achieve a variety of unconventional static and dynamic functionalities. Most two-dimensional configurations involve a single solid layer in which periodicity is induced through voids, as in cellular lattices~\cite{Arretche_Cellular_Grontiers_2018,Vuyk_Elasto-metastructures_EML_2020}, stubs or other surface elements~\cite{Celli-Gonella_Lego-bricks_APL_2015,Jin_Pillars-Moire_JPhysD_2016}, or inclusions~\cite{Lai_Air-inclusion_APL_2003,Lin_Anisotropic-Inclusions_PRB_2011,Matlack_Inclusions_PNAS_2016,Salari-Sharif_Negative-inclusions_PRApp_2019}, or by folding structural components, as in origami and kirigami metamaterials~\cite{Eidini_Origami_EML_2016,Pratapa_Bloch-Origami_JMPS_2018,Rafsanjani-et-al_Kirigami_PNAS_2019}. Fewer concepts have fully embraced the opportunities offered by the interaction of multiple layers. Notable examples include recent works exploring the mechanics of bilayers of pillars~\cite{Jin_Pillars-Moire_JPhysD_2016,Jin_Pillars-Moire_EML_2020}, the interaction of layers of scatterers forming resonant dipoles~\cite{Torrent_Moire_PRApp_2021}, the mechanics of bistable domes~\cite{Udani_Bistable-domes_EML_2021}, and stacked origami structures~\cite{ Zhang_Origami-metastructures_PRE_2020}.
	
	Realizing mechanical analogs of nanoscale bilayers involves the non-trivial task of mimicking their geometric and kinetic characteristics at the macroscale. From a geometry perspective, we need to be able to stack multiple periodic layers. Additionally, the force landscape must entail strong intra-layer forces between neighboring lattice sites within each layer, and weak inter-layer forces between adjacent layers. At the macroscale, the role of the graphene layers can be played by thin hexagonal lattice sheets undergoing flexural deformation. To mimic the weak inter-layer interactions, we can resort to pairs of magnets placed at corresponding lattice sites on adjacent layers. While magnets have been used in phononics applications, either as particles of discrete lattices~\cite{Jiao-Gonella_Interacting-particles_PRB_2020}, or as a tool to endow structural lattices with programmable nodal mass~\cite{vila2017observation} or to promote multistability~\cite{Yasuda-et-al_Transition-Waves_PRApp_2020}, their use as a source of inter-layer interaction is less charted. In principle, an arrangement of two structural layers featuring nodal interaction should fully capture the morphology and kinetics of a graphene bilayer.	\kz{However, a two-layer configuration involving repulsive magnets would be met with a practical limitation. Under the action of the repulsive forces, and in the absence of any confining action, the layers would warp and settle on a deformed configuration that represents the equilibrium point between the magnetic forces and the stiffness of the layers. In these conditions, the remaining magnetic interactions would cease to yield appreciable signature on the layers dynamics.} 
	
\begin{figure} [t]
		\centering
		\includegraphics[scale=0.6]{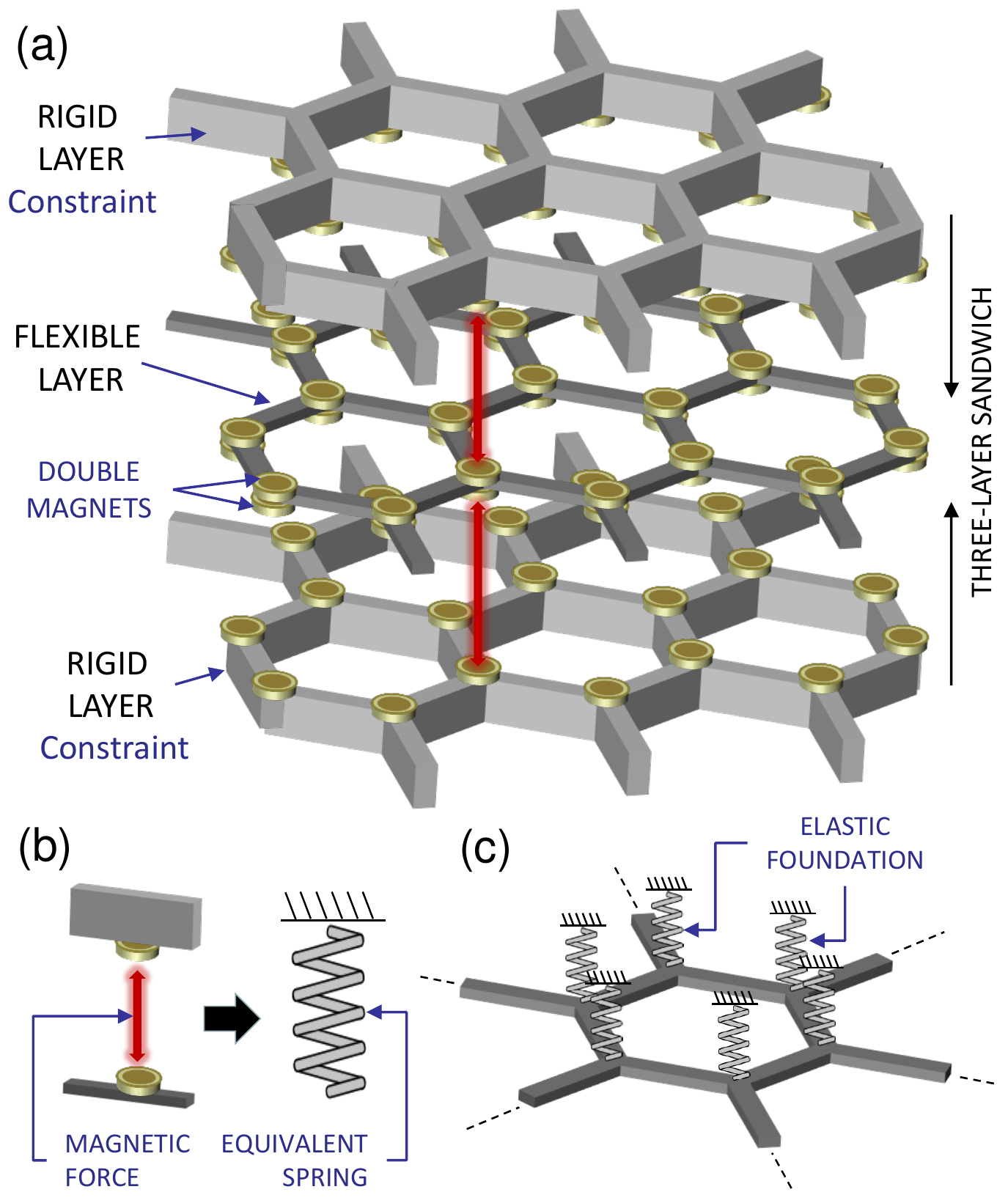}
		\caption{Magneto-mechanical multi-layer analog of graphene bilayer. (a) Three-layer sandwich stacking. (b) Equivalent spring models for the magnetic interactions (c) Inner layer modeled as single layer on elastic foundation.}
		\label{Configuration_schematic}
	\end{figure}

	To overcome this limitation, we consider the three-layer configuration shown in Fig.~\ref{Configuration_schematic}.
	The inner layer, with magnets on both sides, is sandwiched between two outer layers, which also feature magnets on their internal side. The resulting repulsive magnetic forces, exerted symmetrically by the outer layers, keep the inner one in equilibrium. By design, the outer layers are much thicker than the inner one and can therefore be treated, for all intents and purposes, as rigid. As a result, the interactions between pairs of magnets can be modeled as equivalent springs that connect points on the inner layer, which can experience out-of-plane motion, to  fixed points at corresponding locations on the outer layers, thus behaving as an elastic foundation. In essence, even if three layers are practically involved, the system can be effectively modeled as a single honeycomb layer on elastic foundation, with the outer layers serving uniquely a control function.
	
	\begin{figure} [b]
		\centering
		\includegraphics[scale=0.67]{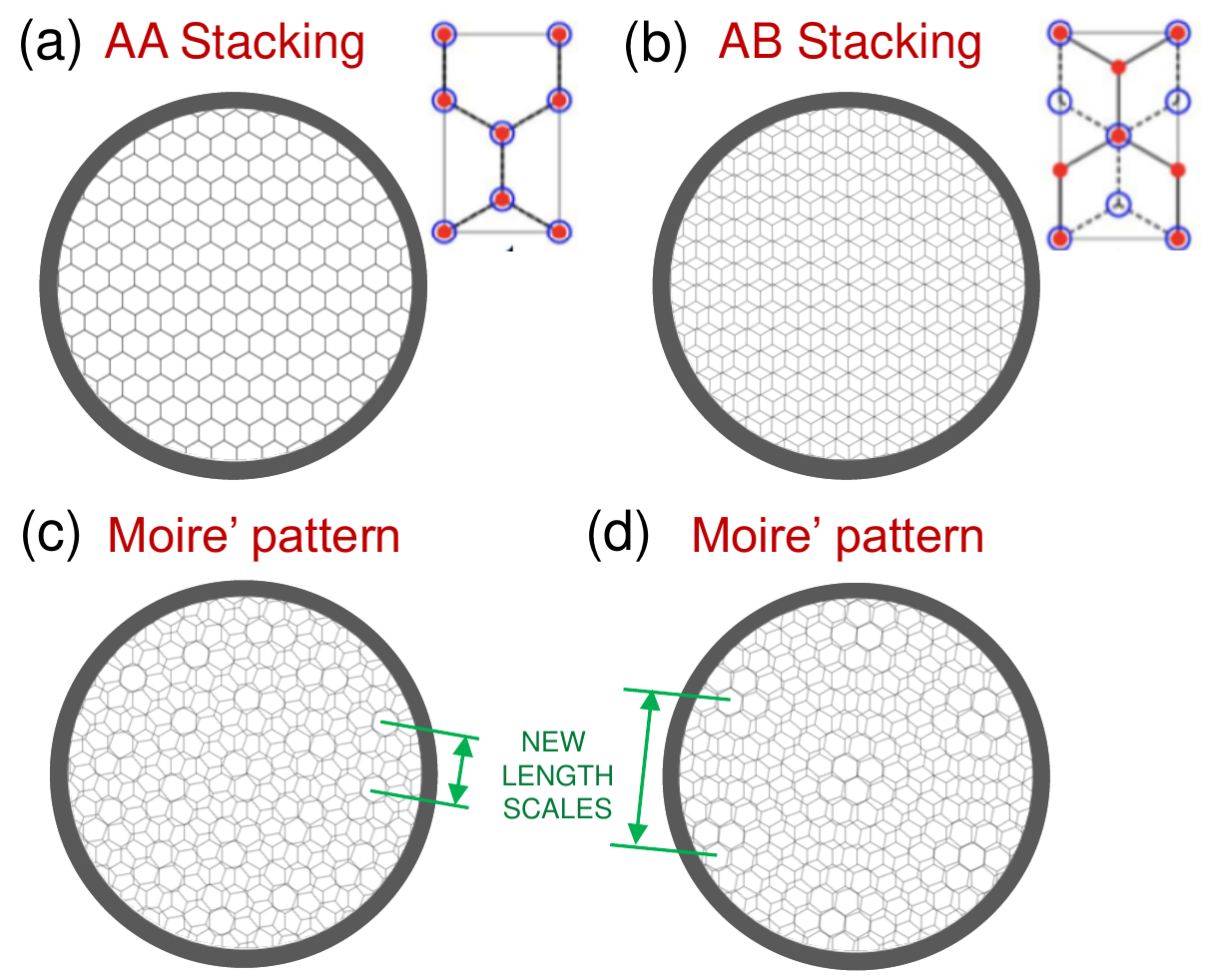}
		\caption{Schematic illustration of the opportunities for reconfiguration available via twisting. (a) AA stacking for twist angle $\theta=0^o$ and perfectly overlapping layers (two lattice sites overlapping per unit cell). (b) AB stacking for $\theta=60^o$ with one lattice site overlapping per unit cell. (c-d) Examples of Moir\'e patterns obtained with intermediate twist angles.}
		\label{Stackings}
	\end{figure}

	It is convenient to shape the layers as circular discs, so that the inner layer can rotate concentrically relatively to the outer ones, resulting in a continuous spectrum of stacking patterns. We also provide the layers with a solid rim through which they can be constrained to one another, upon selection of the desired stacking pattern, and fixed to ground. Four configurations, corresponding to four values of the twist angle $\theta$, are shown in Fig.~\ref{Stackings}. For $\theta=0^o$, the system is in AA stacking (Fig.~\ref{Stackings}(a)), where both lattice sites of each unit cell feel magnetic forces. For $\theta=60^o$, the system is in AB stacking (Fig.~\ref{Stackings}(d)), where only one lattice site per unit cell interacts magnetically. \kz{For intermediate angles, two of which are shown in Fig.s~\ref{Stackings}(b-c), we observe the formation of Moir\'e patterns and the emergence of new length scales in the lattice.} 
	The question is whether different twisted configurations can display different dynamic behavior. Specifically, we are interested in whether the bandgap characteristics of the inner layer can be tuned by twisting. 
	In this Letter we focus on AA and AB stacking, for which we have been able to achieve experimental realizations using our current prototype, as discussed later. 

		\begin{figure} [!htb]
		\centering
		\includegraphics[scale=0.57]{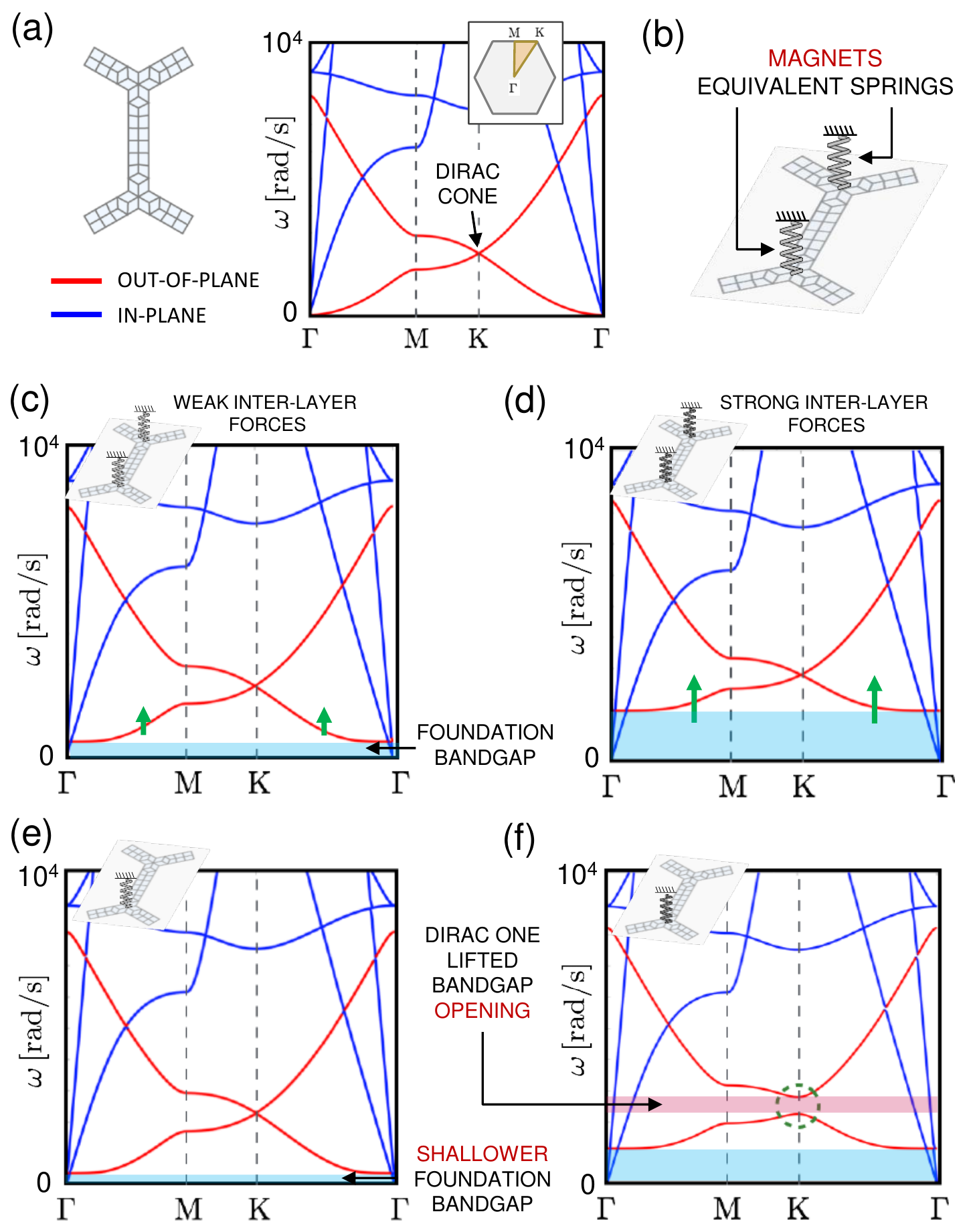}
		\caption{(a) Evolution of band diagram upon changes in twist angle and inter-layer distance. (a) Baseline case of single layer: FEM model of unit cell; band diagram with in-plane (blue) and out-of-plane (red) modes, featuring Dirac cone at $K$ point of the Brillouin zone. (b) Unit cell with elastic foundation (AA stacking case shown). (c) AA stacking with large inter-layer distance (weak magnetic forces), with up-shifting of the flexural mode and opening of a foundation bandgap. (d) AA stacking with smaller inter-layer distance (stronger magnetic forces), resulting in increase of the foundation bandgap width. (e) AB stacking with weak magnetic forces, resulting in reduction of the foundation bandgap width. (f) AB stacking with strong magnetic forces, resulting in increase in foundation bandgap width and opening of finite-frequency bandgap due to space inversion symmetry (SIS) breaking.}
		\label{Band_diagrams}
	\end{figure}

	Using the equivalent spring model, we can effectively explore the effect of different stacking patterns by performing a series of Bloch analyses on unit cells with different foundation spring arrangements. For this illustrative example, we select a set of arbitrary geometric and material properties (whose values are not binding to obtain the results discussed in the following). For material properties, we assume standard values for ABS: Young's modulus $E = 2.14 \times 10^9 \, \textrm{N}/\textrm{m}^2$, Poisson's ratio $\nu = 0.35 $, density $\rho = 1040 \, \textrm{Kg}/\textrm{m}^3$.	The cell beams feature node-to-node length $3.72 \, \textrm{cm}$, in-plane thickness $6.7 \, \textrm{mm}$ and out-of-plane thickness $1 \, \textrm{mm}$. 
    We start from the baseline case of a single layer, whose unit cell 
    is shown in Fig.~\ref{Band_diagrams}(a) with its mesh. The band diagram  of Fig.~\ref{Band_diagrams}(a) features classical in-plane (blue) and out-of-plane (red) modes of hexagonal lattices, with a Dirac cone at the $\textrm{K}$ point, as expected for lattices with six-fold symmetry. 
    
    
	We then proceed to study the effect of AA stacking, by adding to the unit cell two foundation springs, as shown in Fig.~\ref{Band_diagrams}(b). For the equivalent spring constant, here we assume $k_{eq}=10^2 \, \textrm{N/m}$; 
	the characterization of the actual magnetic interaction observed in a physical prototype will be further discussed later in the context of our experimental setup. The corresponding band diagram is given in Fig.~\ref{Band_diagrams}(c), where we see that the elastic foundation lifts the flexural modes, opening a low-frequency foundation bandgap. If we repeat the analysis assuming a larger spring stiffness (here $k_{eq}=10^3 \, \textrm{N/m}$), to model a reduction in inter-layer distance, the width of the foundation bandgap increases, as shown in Fig.~\ref{Band_diagrams}(d). In Fig.s~\ref{Band_diagrams}(e-f), we study the case of AB stacking, modeled by removing one foundation spring. It is important to note that this operation changes the symmetry landscape of the unit cell, relaxing the six-fold symmetry and space inversion symmetry (SIS) while preserving three-fold symmetry. The effects on the band diagram are twofold. For weak magnetic forces (large interlayer distance), the foundation bandgap shrinks, to account for the fact that the cell feels the inter-layer interaction from only one magnet (Fig.~\ref{Band_diagrams}(e)). For stronger magnetic forces (smaller interlayer distance), the signature of symmetry breaking becomes distinct, with the lifting of the degeneracy at the Dirac point and the opening of a finite-frequency bandgap. We note that the SIS relaxation is one of the ingredients required to design mechanical quantum valley Hall effect (QVHE) analogs used to realize non-trivial waveguides~\cite{vila2017observation,zhu2018design, Ma-et-al_Valley-Hall_PRApp_2019}. Interestingly, while in most QVHE analogs the symmetry is broken through physical changes in the cell, here the effect is obtained in a contact-less fashion through the interaction with the control layers, and is therefore fully reversible.

	\begin{figure*} [!htb]
	\raggedright
	\includegraphics[scale=0.7]{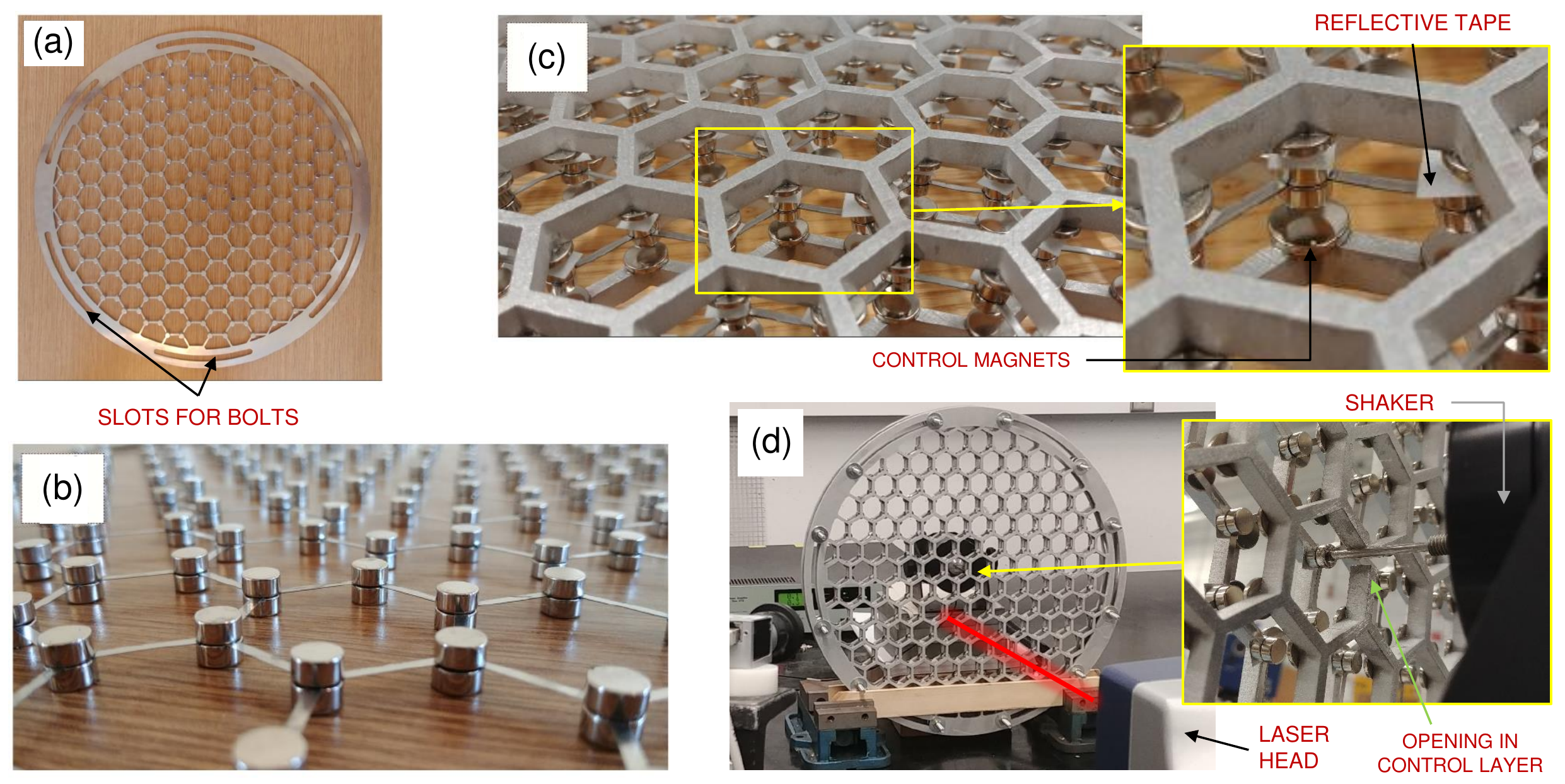}
	\caption{Experimental prototype and setup. (a) Inner layer with solid rim and curved slots for bolts allowing for adjustable twist. (b) Detail of the inner stainless steel layer, showing double magnet arrangement and thin ligaments. (c) Detail of the tri-layer sandwich structure showing co-axial interaction between pairs of magnets. (d) Setup showing base clamp, point excitation provided via electromechanical shaker and scanning laser Doppler vibrometry (SLDV) scan.}
	\label{Setup}
	\end{figure*}
	
	To test the tuning capabilities, we designed and assembled the prototype shown in Fig.~\ref{Setup}. 
	The three layers are cellular discs with identical diameter and lattice constant, and a solid rim with overlapping slots designed to enable fastening while allowing for adjustable twist. \wj{The inner layer, shown in isolation in Fig.~\ref{Setup}(a), and the outer layers are water-jet cut from a $0.15 \, \textrm{mm}$-thick sheet of stainless steel (type 304, with Young's modulus $E = 190 \times 10^9 \, \textrm{N}/\textrm{m}^2$, Poisson's ratio $\nu = 0.265 $, density $\rho = 7850 \, \textrm{Kg}/\textrm{m}^3$) and from a $6 \, \textrm{mm}$-thick slab of aluminum (treated as rigid), respectively. We glue Neodymium N-42 permanent magnets on both sides of the inner layer, as shown in Fig.~\ref{Setup}(b), and on one side of each outer layer, such that opposing magnets on adjacent layers have opposite polarizations. 
	The magnets on the inner and outer layer have diameters $1/4 \, \textrm{in}$ and $3/8 \, \textrm{in}$, and thicknesses $1/8 \, \textrm{in}$ and $1/16 \, \textrm{in}$, respectively}. Finally, we stack the lattices as to establish repulsive magnetic interactions between each layer pair, as shown in Fig.~\ref{Setup}(c). The layers are separated by washers and fastened 
	through the rim slots. Thus, the strength of the magnetic interaction can be tuned in discrete increments by controlling the spacing (number of washers) between the layers. 
	
	By rotating the inner layer, we can establish different stacking patterns. 
	\wj{Although all twist angles can be achieved in principle, in this prototype only $\theta=0^o$ and $\theta=60^o$, corresponding to AA and AB stacking, respectively, lead to stable configurations. This is a side effect of the magnets, which are repulsive axially but attract each other side-by-side. For small twist angles, some magnet pairs are displaced such that they start feeling strong in-plane attraction despite living on different layers - a phenomenon that is accentuated at small inter-layer distances. The in-plane forces may occasionally result in magnets being stripped from the lattice; they can force the ligaments of the inner layer to warp macroscopically, altering the layer shape; or they can cause the layers to come into contact, ultimately collapsing the structure.} For these reasons, here we focus on the dichotomy between the extreme stacking scenarios AA and AB. 
	
	\begin{figure*} [t]
	\raggedright
	\includegraphics[scale=0.7]{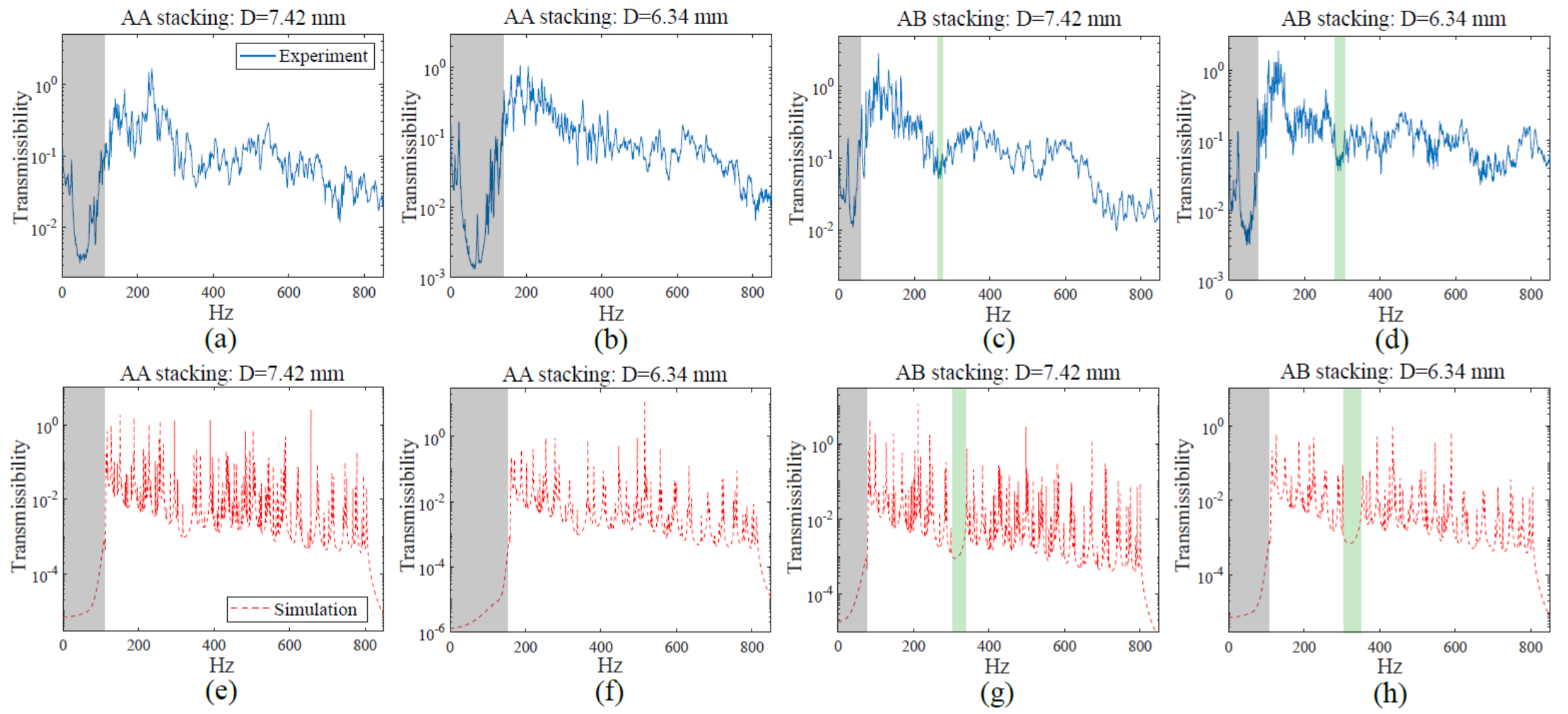}
	\caption{Experimental results. (a-d) Transmissibility vs. frequency plots obtained from vibrometry data for AA stacking (a-b) and AB stacking (c-d), for different inter-layer distances. The plots revel a foundation bandgap, whose width depends on the inter-layer distance and can tuned by twist. For sufficiently strong inter-layer interactions, a second bandgap associated with symmetry breaking is switched on by twisting the layers from AA to AB stacking. (e-h) Transmissibility plots obtained from full-scale simulations for same stacking conditions and inter-layer distances, correlating well with the experimental results.}
	\label{FRF_Results}
	\end{figure*}

The experimental setup is shown in Fig.~\ref{Setup}(d). A Scanning Laser Doppler Vibrometer (SLDV, Polytec PSV-400-1D) is used to measure the out-of-plane response of the inner layer. The specimen is placed vertically and clamped at the bottom. The out-of-plane excitation is imparted at the center of the inner lattice through an electrodynamic shaker (Bruel $\&$ Kjaer Type 4809) powered by a Bruel $\&$ Kjaer Type 2718 amplifier. Note that the shaker stinger engages the inner layer through an opening in the rear outer layer, see inset of Fig.~\ref{Setup}(d). Using this setup, we excite harmonically, sweeping the frequency of excitation. For each frequency, we measure with the vibrometer the velocity of a lattice site, implementing several averages to filter out noise. 
We acquire and average velocity values from multiple scan points to obtain a single scalar output measure. Finally, we construct a transmissibility metric by dividing this average output velocity by the velocity of the excitation point. 

The transmissibility is plotted against frequency in Fig.~\ref{FRF_Results}(a-d). Figures~\ref{FRF_Results}(a-b) refer to AA stacking, for inter-layer distance $\textrm{D}=7.42 \, \textrm{mm}$ and $\textrm{D}=6.34 \, \textrm{mm}$, respectively. We observe a large foundation bandgap, whose width increases as the inter-layer distance decreases, in accordance to the model. 
Surprisingly, we note that the bandgap does not start at 0 $\textrm{Hz}$. \wj{While a combination of factors probably concur to produce this deviation from the model, the most likely explanation lies in the non-ideality of the boundary constraints in the experimental setup. A bandgap starting at 0 $\textrm{Hz}$ requires a perfect foundation, with springs connecting the structure to fixed points. While the tri-layer sandwich is held in place by firm clamps, the possibility for these clamps to slide slightly on the table surface (especially at low frequencies where the imposed displacements are large) makes the physical realization of the foundation imperfect. The peaks 
below the bandgap are likely to capture these very low-frequency 
vibrational modes.} 
The results for AB stacking, for the same pair of inter-layer distances, are shown in Fig.s~\ref{FRF_Results}(c-d). The first thing to notice is the conspicuous shrinking of the foundation gap, in accordance to the theory. The second effect is the opening of a second bandgap at finite frequencies (here around 270 $\textrm{Hz}$). This effect is pale for $\textrm{D}=7.42 \, \textrm{mm}$ in Fig.~\ref{FRF_Results}(c) but becomes evident for  $\textrm{D}=6.34 \, \textrm{mm}$ in Fig.~\ref{FRF_Results}(d). 

We accompany the experiments with full-scale FEM simulations of the entire lattice disc, shown in Fig.s~\ref{FRF_Results}(e-h). 
\wj{To estimate the equivalent stiffness of the foundation springs, one could rely on manufacturers and  charts. However, these are only accurate if the magnets are perfectly co-axial, and are very sensitive to the inter-magnet distance. In the experiment, even small warping of the inner layer results in partially tilted magnets that lose co-axiality 
and in a non-uniform landscape of distances.} For these reasons, we prefer to leave the equivalent magnets stiffness as a free calibration parameter. The calibration is performed such that the width of the foundation bandgap for AA stacking matches the one observed in the experiments, as shown in Fig.s~\ref{FRF_Results}(e-f). The predictive quality of the calibrated model can then be tested by comparing the frequency responses for AB stacking in Fig.s~\ref{FRF_Results}(g-h) against their experimental counterparts. The shrinking of the foundation gap is well captured. The emergence of a finite-frequency gap due to symmetry breaking is also predicted, although the location appears slightly upshifted compared to the experiments.
	
In conclusion, we have demonstrated experimentally that we can tune the bandgap response of a magneto-mechanical multi-layer metastructure by controlling the relative twist between the layers. 
While, in this work, we have focused on 
AA and AB stacking, the analysis can be conceptually extended to intermediate configurations corresponding to a full spectrum of Moir\'e patterns. 

This work was supported primarily by the National Science Foundation through the University of Minnesota MRSEC under Award Number DMR-2011401. The authors also acknowledge partial support from the National Science Foundation (CAREER Award CMMI-1452488). 
	
	\bibliography{bib}
				
\end{document}